\documentclass[a4paper,11pt]{article}
\usepackage{a4}
\usepackage{amsmath}
\usepackage{amsfonts}
\usepackage{graphicx}
\DeclareGraphicsExtensions{.eps,.eps.gz,.eps.Z,.jpg,.pdf,.bmp}
\usepackage{times}
\usepackage[T1]{fontenc}
\usepackage{indentfirst}

\title
{\bf{ELTs Adaptive Optics for Multi-Objects 3D Spectroscopy : Key Parameters and Design Rules}} 
\author
{B. Neichel$^{a,b}$, J-M. Conan$^b$, T. Fusco$^b$, E. Gendron$^c$, M. Puech$^a$,  \\G. Rousset$^a,b$ and F. Hammer$^a$}
\date{}

\begin{document}
\maketitle

{\small
\begin{center}
$^a$GEPI, Observatoire de Paris, 5 place Jules Janssen, 92195 Meudon, France.\\
email : benoit.neichel@obspm.fr \\
$^b$ONERA, BP 72, 92322 Chatillon Cedex, France\\
$^c$LESIA, Observatoire de Paris, 5 place Jules Janssen, 92195 Meudon, France. 
\end{center}}

\begin{abstract}
In the last few years, new Adaptive Optics [AO] techniques have emerged to answer new astronomical challenges: Ground-Layer AO [GLAO] and Multi-Conjugate AO [MCAO] to access a wider Field of View [FoV], Multi-Object AO [MOAO] for
the simultaneous observation of several faint galaxies, eXtreme AO [XAO] for
the detection of faint companions. In this paper, we focus our study to one of these applications : high red-shift
galaxy observations using MOAO techniques in the framework of Extremely Large Telescopes [ELTs]. 
We present the high-level specifications of a dedicated instrument. We choose to describe the scientific requirements with the following criteria : $40$\% of Ensquared Energy [EE] in $H$ band ($1.65\mu$m) and in an aperture size from $25$ to $150$ mas. Considering these specifications we investigate different AO solutions thanks to Fourier based simulations. Sky Coverage [SC] is computed for Natural and Laser Guide Stars [NGS, LGS] systems. We show that specifications are met for NGS-based systems at the cost of an extremely low SC. For the LGS approach, the option of low order correction with a faint NGS is discussed. We demonstrate that, this last solution allows the scientific requirements to be met together with a quasi full SC.

\end{abstract}


\section{INTRODUCTION}
\label{sect:intro}  
A new era of astronomical telescopes is about to start with diameters
reaching $30$ to $60$ meters.
This concept of ELTs will
provide a dramatic advance in our understanding of the primordial universe.
By accommodating 3D
spectroscopy devices and relatively large FoV ($10$ arc minutes),
ELTs will be a privilaged tool for the study of formation and evolution
of galaxies.\\
The spatial resolution of ground based telescopes is limited by the
presence of the atmospheric turbulence which leaves us with image resolution of $0.5$ arc second at best. 
Current Very Large Telescopes [VLTs]
($8$-$10$ meter-class) are now equipped with AO which provides a
real time correction of turbulence and enables large telescopes to give almost diffraction
limited images. But because of anisoplanatism effects, the AO corrected FoV is small, and a good correction requires bright (Natural or Laser\cite{Foy85}) Guide Stars near the scientific object. Typically, the compensated FoV around the Guide Star is of the order of a few times the isoplanatic patch $\theta$$_0$ (a few tens of arcsec in the Near InfraRed).
In extragalactic studies, to avoid contamination of light by the
interstellar medium, it is necessary to observe in a direction far of our
galactic plane. In that case, the surface density of stars becomes very small,
and classical AO working with NGS can not be used. Even close to the galactic plane, because of  the required star brightness, SC would be lower than $5$\%.\\
To overcome this problem, new AO techniques (GLAO\cite{Hubin05}, MOAO\cite{Hammer01,Gendron05})
have emerged in the last few years to increase the corrected
field. These new methods are based on a full measurement of the 3D turbulent volume using several Guide Star.
The correction is then applied using one or more Deformable Mirrors [DMs] and the corrected FoV can reach few arcmins. To improve the sky coverage, LGS have been
proposed\cite{LeLouarn98}. This solution, that create an artificial star in the sodium layer of the atmosphere, could be used where the density of stars becomes too small for a good turbulence analysis even with GLAO or MOAO.
Should some critical issues be solved (cone effect,
spot elongation, tilt indetermination), the gain achievable by such a system could be very promising. The ultimate solution would be to associate LGS with multi-analysis concepts\cite{Lelouarn02} .
\\
These instruments are now under development
for VLTs. The AO design rules and choices to meet a given performance
are relatively well understood for this kind of telescope diameter. 
However, for ELTs the whole instrumentation has to be reconsidered and redefined. \\
\\
In section $2$ we present the instrumental specifications imposed by astrophysical goals. Section $3$ is dedicated to a preliminary estimation of an AO design and the SC issue is discussed. In Section $4$ we analyze the performance of an AO system using laser probe, and in Section $5$ we extend this study to Wide Field AO concepts.

\section{High Level Specifications} 
 
Extragalactic studies will benefit from the large
capabilities of ELTs in light concentration
and spatial resolution. 3D Spectroscopy of galaxies up to $m_{AB}$ $\simeq$ $25$ will have a large impact in our understanding of the assembly of dark and
visible matter (from $z=0$ to $z=5$), the physics of galaxies near the
reionisation ($z=6$-$9$) and the search for the primordial galaxies ($z>>6$)\cite{Hook04} .
\\
\\
3D Spectroscopy of distant galaxies will necessitate to
obtain spatially resolved spectra of very faint and physically small
objects.
In terms of instrumental specifications this leads to several critical points.

\subsection{Spectral Resolution} 
\begin{figure}[h!]
   \begin{center}
   \begin{tabular}{c}
   \includegraphics[width = 0.4 \linewidth]{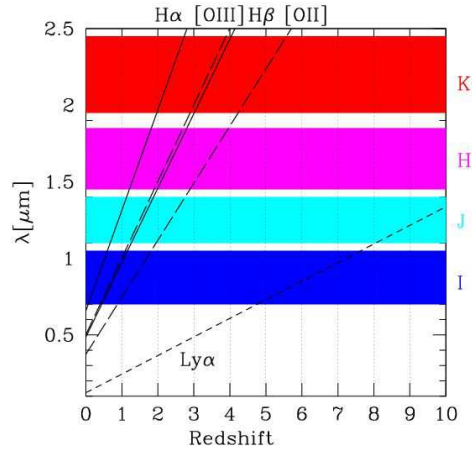}
   \end{tabular}
   \end{center}
   \caption[example] 
   { \label{fig1} 
Spectral localization of emission lines at different redshifts from $I$ band to $K$ band (from $0.7\mu$m to $2.45\mu$m respectively). We see that NIR has a special importance for the study of emission lines at high redshift.}
   \end{figure}

Figure \ref{fig1} shows that the most important emission lines in galaxy studies ($H_{\alpha}$, $H_{\beta}$, [OII], [OIII]) will be redshifted in the Near InfraRed [NIR] at high redshift. 

In this spectral window, the sky is dominated by thin, but intense OH emission lines\cite{Rousselot00} . To avoid contamination by these atmospheric OH lines, a minimal spectral resolution is needed to separate them and make observations between these lines possible. That leads to a minimal spectral resolution of $R>5000$. \\
Besides, the scientific goal for the study of galaxy dynamics would be to obtain resolved velocities of a few tenths of kilometers. To achieve this at high redshift ($z\sim5$), a spectral resolution of $R = 10000\pm5000$ is required.
\newpage

\subsection{Spatial Resolution}
\begin{figure}[h!]
   \begin{center}
   \begin{tabular}{cc}
   \includegraphics[height=5cm]{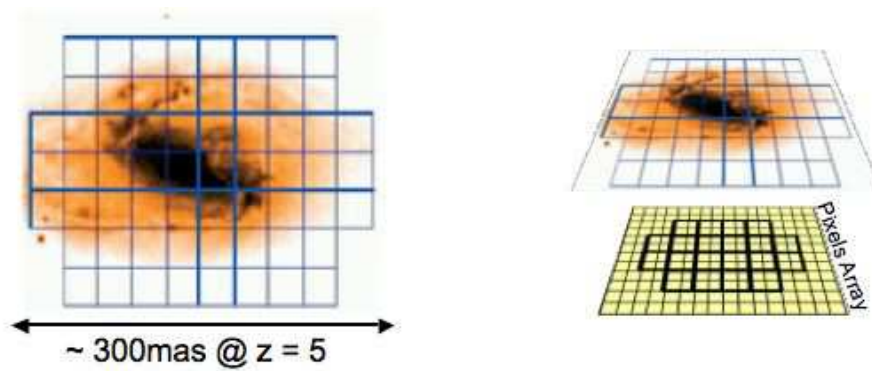}
   \end{tabular}
   \end{center}
   \caption[example] 
   { \label{aperture} 
\textit{Left} : Illustration of the spatial sampling of a galaxy. Each square represents a spatially resolved resolution element in which a spectra is obtained. \textit{Right} : To spatially sample each resolution element, the pixel size is twice smaller.}
   \end{figure}

The typical size of galaxies at $z\simeq5$ is $R_{half}\simeq150$mas (half light radius 
in the stellar continuum\cite{Bouwens04}). Sampling $R_{half}$ would then require an angular resolution of 
at least $75$ mas. To map a physical parameter on these objects, and thus be able to resolve different areas (HII regions, SN, ...) an even better spatial resolution is required (resolution element $\sim25$mas - see figure \ref{aperture} on the left). Note that, based on Shanon crtiterion, two pixels per resolution element are needed (see figure \ref{aperture} on the right).
The angular resolution specification can probably be relaxed as first GIRAFFE results\cite{Puech05} (3D spectrograph at the ESO-VLT) have shown that in many 
distant galaxies, the gas seems to be more extended than the stars. An angular resolution around $R_{half}$ could probably be sufficient. In the present study, we choose to explore an angular resolution size ranging 
from $25$ to $150$ mas.

\subsection{Field of View and multiplex capability}
A large FoV will be required to avoid cosmic variance effects, and obtain a
statistically unbiased sample. A large FoV is also essential to allow the simultaneous 3D spectroscopy of several galaxies. To encompass projected clustering scales at high redshift (which size varies from 4 to 9 Mpc), a field of view of $10$ arcmin in diameter is necessary. A first rough estimation of the instrument multiplex capability can be estimated by considering the number density of 
galaxies at $z\simeq5$, where the most emission lines leave the $K$ band. Lehnert et al.\cite{Lehnert03} give a number density of $\sim$ 0.3 source per square arcminute between $z=4.8$ and $z=5.8$ down to $I_{AB}\simeq26$. A $10$ arcmin diameter FoV would then necessitate a minimal multiplex capability of few tens of objects.

\subsection{Ensquared Energy [EE]}
We define a parameter to link the image quality with the spectroscopic performance. This parameter is the fraction of Energy of a PSF, normalized to the total flux, Ensquared within a given square aperture. In our case, the Useful Aperture [UA] is a square box which size fits the specified angular resolution (from $25$ to $150$ mas). 
Assemat et al.\cite{Assemat06} show that, for a $8$-meter class telescope, an EE of $30$-$40$\% is required to
detect the H$\alpha$ line with a Signal to Noise Ratio [SNR] greater than $3$.
 More recently, preliminary results (Puech et al., in preparation) based on simulations of  kinematics of distant galaxies, show that physical parameters are recovered only with an EE of at least $30$\% for a $40$-meter class telescope.\\
We see that both results are consistent, and we choose to set this scientific specification to
40\% of EE in $H$ band in the following study. This could be subject of future adjustments, but seems
to be a reasonable work baseline.

\subsection{Seeing limited performance}
\begin{figure}[h!]
   \begin{center}
   \begin{tabular}{c}
   \includegraphics[height=7cm]{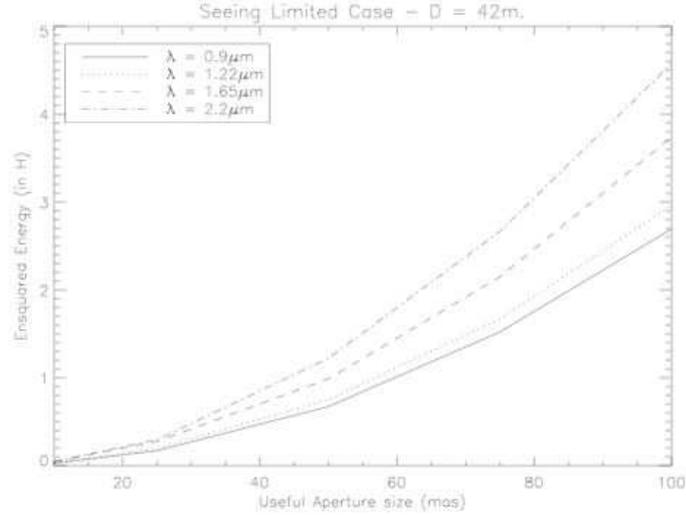}
   \end{tabular}
   \end{center}
   \caption[example] 
   { \label{fig2} 
Percentage of Ensquarred Energy in a growing UA for a seeing limited instrument. Seeing = $0.81$arcsec. $\lambda$ is the scientific wavelength. Note the small Y axis range values.}
   \end{figure}

Figure \ref{fig2} shows the expected EE in different UA sizes for an instrument limited by the atmospheric turbulence. We see that with a seeing-limited instrument, the performance reached is very far from the scientific specifications.

The size of the UA is typically ten times smaller than the seeing-limited image. AO is thus essential to concentrate the light within these small apertures. AO will also increase the SNR as it minimizes the flux spread in the neighboring apertures.

\subsection{Multi-Objects Adaptive Optics}
Taking into account the huge FoV determined by scientific considerations, only Wide Field AO systems could be used for Extragalactic Studies. In a previous study\cite{Neichel05}, we have shown that a GLAO system is not consistent with the foreseen performance. We will thus focus on Multi-Objects Adaptive Optics systems.\\
Instead of compensating the whole field, MOAO will perform the correction locally on each scientific object. Several off axis Guide Stars are considered to perform a tomographic measurement of all the turbulent volume around each scientific
object (ie. direction of interest). The optimal correction is then deduced from the turbulence volume knowledge and applied
using a single DM per direction of interest.\\
\newpage
In order to design our MOAO system, we assume that the related residual phase variance can be split as :
\begin{equation}
\label{budget}
\sigma_{res}^2=\sigma_{fit}^2+\sigma_{alias}^2+\sigma_{temp}^2+\sigma_{noises}^2+\sigma_{tomo}^2
\end{equation}
The first term in eq. \ref{budget} represents the undermodelling error, the second term is the Wave Front Sensor [WFS] aliasing error, the third term is the servo-lag error, the fourth term corresponds to the WFS noise (photon and detector) and the last term is the error link with tomographic reconstruction of the turbulent volume.\\
\\
In the following, we will study how to distribute the weight conceded for each error term, in order to reach the EE specification. 

\section{The Natural Guide Star solution} 
\label{AOonAxis}
In this section, we are interested on how a "classical" NGS-based AO system should be designed (number of actuators, GS magnitude, ...) in order to reach the scientific specifications. Instead of dealing with multi-NGS tomographic AO right now, we want first to consider the simple on-axis case. This will give us an upper limit value of the performance, knowing that in practice, multi-NGS systems performing tomography reconstruction should lead to supplementary terms of errors (e.g. tomographic error). These terms will be analyzed in more detail later on. The goal of this section is not to give a detailed analysis of the system but rather to find general behaviors.
\subsection{Dimensioning the AO system}
To perform our simulations, we used a simulation tool based on Fourier approach which computes an AO corrected PSF for different AO configurations\cite{Conan03,Jolissaint06} . We study the EE versus several parameters : the size of the UA, the correction degree, the temporal bandwidth and the magnitude of guide stars.
\subsubsection{Degree of correction}
\label{doc}
\begin{figure}[h!]
   \begin{center}
   \begin{tabular}{cc}
   \includegraphics[height=7cm]{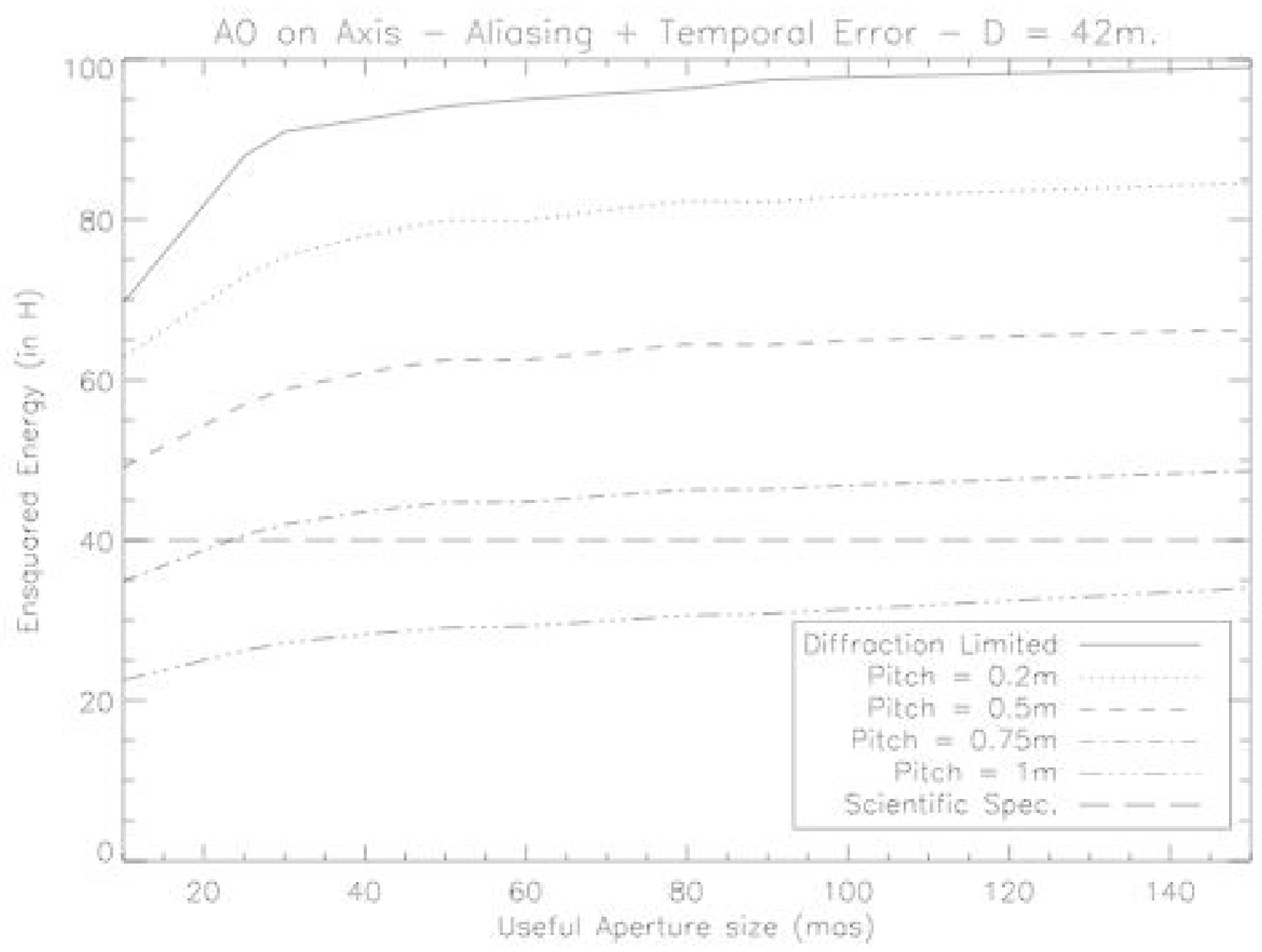} &
         \includegraphics[height=7cm]{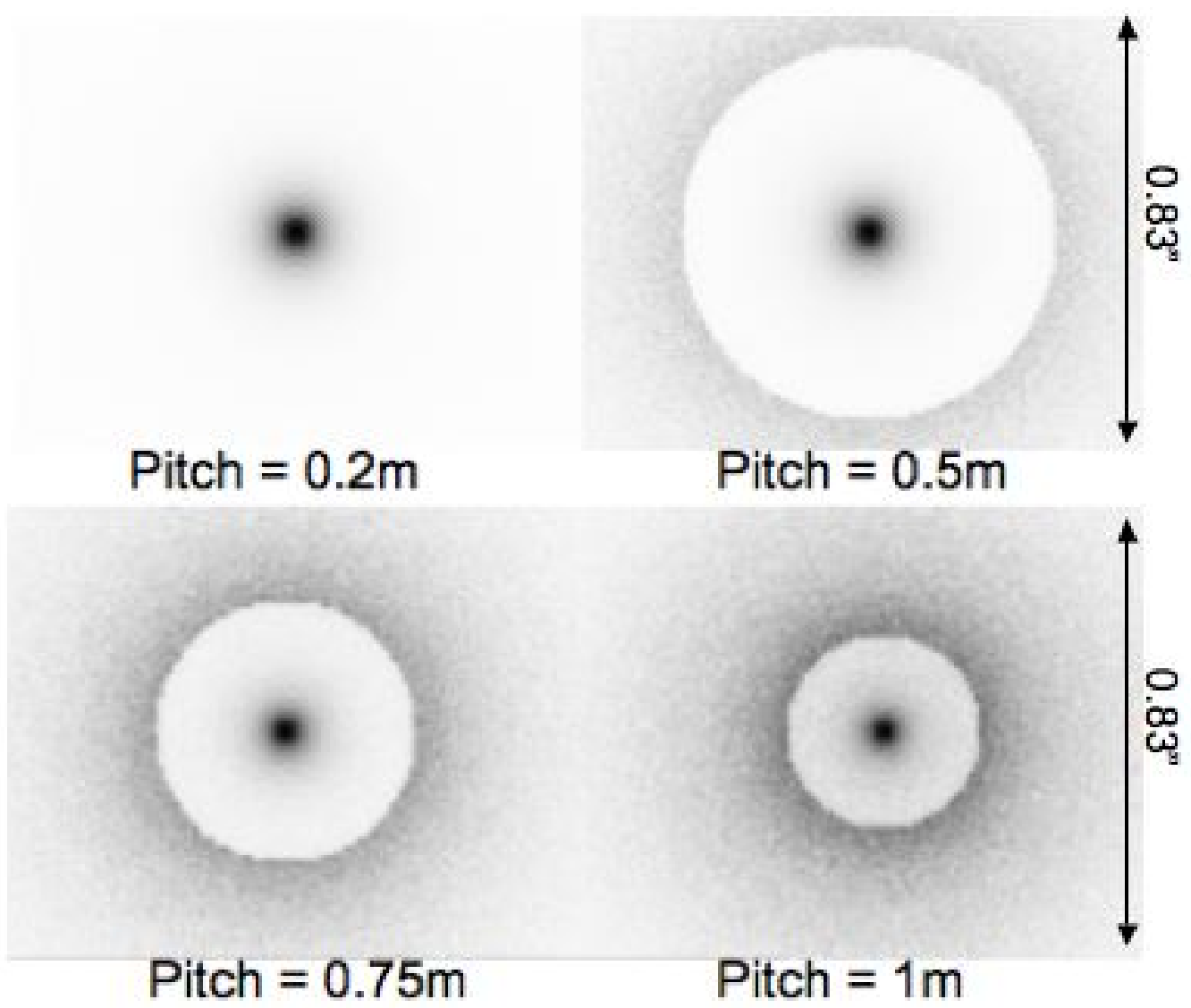}
   \end{tabular}
   \end{center}
   \caption[example] 
   { \label{fig3} \textit{Left} : Effect of the size of the UA on the expected performance for various degrees of correction. \textit{Right} : AO-corrected PSFs for different pitch sizes (log scale).}
   \end{figure}

The first step to dimension our AO system is to evaluate how many actuators (or WFS sub-apertures) are needed to concentrate, closer to the center, the energy spread far from the optical axis by the turbulence. We consider undermodelling, aliasing and servo-lag errors, that are closely related to correction degree, and we neglect WFS noises and anisoplanatism. For undermodelling, we assume that the AO correction is characterized by the DM cut-off frequency which is defined by :
\begin{equation}
f_c=\frac{1}{2d} 
\end{equation}
where $d$ (hereafter called pitch) is the WFS sub-aperture size. \\
\\
The number of actuators (or WFS sub-apertures) is then defined by :
\begin{equation}
N=\frac{D}{d}
\end{equation}
where D is the telescope diameter.\\
\\
For Servo-lag error, we assume that the turbulence is stratified in $10$ independent horizontal layers. Each layer is seen as a frozen screen propagating horizontally at constant velocity and direction across the telescope pupil. The temporal sampling frequency is then optimized to maximize the EE. Finally, the aliasing error results from the spectral aliasing of the high-order modes into lower order modes.\\
The main atmospheric parameters and the AO hardware characteristics are summarized in Table \ref{tab1}. Simulation results are presented in Figure \ref{fig3}.\\
\\

\begin{table}[h!]
\begin{center}
\begin{tabular}{|c|c|c|}
\hline \bf{Parameter} & \bf{Value} & \bf{Remark}\\ \hline \hline
Diameter & $42$m & No central obstruction \\ \hline
Turbulence & Typical Paranal Profile &  10 layers \\ \hline
Seeing & $0.81"$ @ $0.5$ $\mu$m &  Paranal median value \\ \hline
$L_0$ & $22$m & Paranal median value \\ \hline
$<$Wind$>$ & $12.5$ m/s & \\ \hline
Wave Front Sensor & Shack-Hartmann @ $0.65$$\mu$m & Same number of lenslets \\ 
type & $4$x$4$ pixels/sub pupils & as actuators \\ \hline
WFS Temporal Sampling & $50$-$500$Hz & Total delay = $2$ frames \\ \hline 
\end{tabular}
\caption{\label{tab1} Simulation parameters.}
\end{center}
\end{table}

\newpage
   Figure \ref{fig3} shows that :
\begin{itemize}
\item The size of the UA has a small impact on the performance as soon as this size is large enough ($>30$-$40$mas) compared to the diffraction limit. Indeed, as shown in Figure \ref{fig3} (right), the PSFs are composed by a diffraction core which size is small ($\lambda$/D$ \simeq8$mas) and a large residual halo ($\lambda$/pitch $\simeq680$mas for pitch = $0.5$m). As soon as the size of the UA is comprised between these particular sizes, the EE will nearly remain constant. Of course this is only valid for this kind of PSF shape.
\item To achieve the performance imposed by scientific specifications, a large number of actuators is needed. Even for this optimistic case of on-axis correction with bright NGS, a pitch size smaller than $0.75$m is required. For a $40$m-class telescope, that leads to at least a $50$x$50$ actuators. A high coupling factor can only be obtain by using a high order compensation.
\end{itemize} 

\subsubsection{Impact of GS magnitude}
\label{mag}
\begin{figure}[h!]
   \begin{center}
   \begin{tabular}{c}
   \includegraphics[height=7cm]{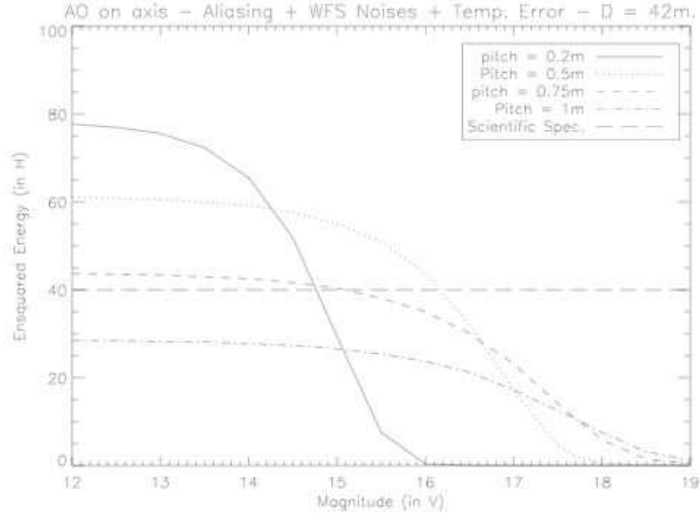}
   \end{tabular}
   \end{center}
   \caption[example] 
   { \label{fig4} Impact of the magnitude of guide star on the EE versus the degree of correction. The UA size is set to $50$mas in this example. Magnitude are given in V band.\\}
   \end{figure}

Multiplying the number of actuators (or WFS sub-apertures) constraints the limiting magnitude of stars available for WFS. To illustrate this limitation, we compute the impact of the magnitude of the guide star by adding WFS noises (photon and CCD) to the previous simulations. For the WFS photon noise, we assume a global transmission of our system given by a Zero Point = $10^{10}photo$-$e^-/m^2/s$. This Zero Point is defined by the number of photo-electrons detected on the WFS CCD for a G0 star with V = 0. It includes the overall transmission of atmosphere and telescope optics ($\tau = 0.25$), the quantum efficiency of the detector ($\eta = 0.9$) and the spectral bandpass of the CCD ($\Delta_{\lambda} = 0.4\mu$m). For the electronic noise, we consider a CCD providing $1$e-/pixel. Note that no modal optimization was taken into account in these simulations. Figure \ref{fig4} shows the impact of these additive noises on the expected performance for different degrees of correction.

\newpage
Figure \ref{fig4} shows that relatively bright stars must be considered for WFS as the received number of photon is divided between a large number of sub-pupils. For instance, we note that for a pitch value around $0.5$m (resp. $0.2$m), only stars with magnitude lower than $16.2$ (resp. $14.8$) could be used for WFS in order to reach $40$\% of EE. The best compromise between performance and limiting magnitude seems to work with a $0.5$m pitch system ($\sim80$x$80$ actuators). In the following study we will set the pitch size to this value.

\subsection{Sky Coverage issue}
Results from figure \ref{fig4} point out that a limitation in the choice of the NGS magnitude appears due to the high degree of correction needed. When dealing with NGS, this limitation will lead to SC issue. \\
The purpose here is to estimate the SC when working with NGS in a tomographic fashion. In a MOAO system, several off axis GS (a constellation), even far outside the isoplanatic patch, are used to perform a
tomographic measurement of all the turbulence volume. Depending on the size of the constellation and the magnitude of the stars, the achievable SC will change. Unfortunately, for widely separated GS, the tomographic error can be very high. For this reason, we choose to investigate reasonable separated constellations where reference stars are sought within a $2.5$ or $5$ arcmin FoV in diameter. We also presumed that the limiting magnitude for three NGS will be approximately the same as for one NGS. From figure \ref{fig4} we choose V$\le16$. For all our SC estimation we have used an algorithm developed by Fusco et al.\cite{Fusco06}
Sky coverage results are presented in Table \ref{result}.
\begin{table}[h!]
\begin{center}
\begin{tabular}{|c|c|c|c|c|c|c|}
\cline{2-7}
\multicolumn{1}{c|}{}&  \multicolumn{2}{c|}{\bf{Galactic Latitude = $30^{\circ}$}} & \multicolumn{2}{c|}{\bf{Galactic Latitude = $60^{\circ}$}} & \multicolumn{2}{c|}{\bf{Galactic Pole}}\\ \cline{2-7} \hline
\textit{V $\le 16$} & FoV = $2.5'$ & Fov = $5'$& FoV = $2.5'$ & Fov = $5'$& FoV = $2.5'$ & Fov = $5'$ \\ \hline
Tomographic Mode& 7 \%& 40.3\%  & 0.5\% & 8.5\% & 0.3\%  & 5.3\%  \\ \hline\hline
Single NGS Mode &  \multicolumn{2}{c|}{4.6\%} & \multicolumn{2}{c|}{1.7\%} &  \multicolumn{2}{c|}{1.3\%}\\ \hline
\end{tabular}
\caption{\label{result} Sky Coverage for different galactic latitudes and V $\le 16$. In the Tomographic Mode, the SC is expressed in the percentage of the sky where at least 3 stars with $V \le 16$ can be found in a given FoV. In the Single NGS Mode, the SC is given by the number of stars with $V \le 16$ multiplied by the isoplanatic surface.}
\end{center}
\end{table}

Even when working with NGS in a tomographic configuration, full SC can be obtained only near the Galactic plane. This is incompatible with extragalactic studies : they require to observe at high galactic latitudes, where the Sky Coverage becomes small. To overcome this limitation, the solution of using an Laser Guide Star is mandatory. Only the LGS solution could offer a sufficiently important SC, even near the galactic pole. In the following study we will focus on systems working with LGS.

\section{The Laser Guide Star Solution}
\label{llgs}
LGS systems are usually based on the analysis of the back scattered light from the excitation of an atmospheric resonant layer. This solution, which can provide bright sources anywhere in the sky seems very attractive. Nevertheless, this promising technique suffers from some limitations detailed below.\\
A first problem is due to the finite altitude of the LGS. Rays from the LGS do not sample the same turbulence as the collimated beam coming from the science object leading to a phase estimation error. This effect (which is called Focus Anisoplanatism [FA] or Cone effect) becomes predominant for ELTs, and makes the use of a single LGS useless. However, previous studies\cite{TF90,Viard02} have shown that FA can be solved by using several LGS to sense the whole cylinder turbulence path. In the following we will make the assumption that FA error is actually solved.\\
A more severe limitation is the Tilt indetermination. As a consequence of the round trip of light the wavefront tilt cannot be obtained from the LGS\cite{Pilkington87}. Elaborate techniques have been proposed to measure the Tilt from the LGS\cite{Foy95,Ragazzoni96} but unfortunately, real time correction has not yet been demonstrated. This forces to use a NGS to retrieved the wave front Tilt\cite{Rigaut92}.\\
In the following sections, we will consider an ideal system which provides a perfect correction of modes higher than Tilt. All other terms of errors are neglected. The purpose is no more to achieve a given scientific specification, but rather to evaluate the impact of Tilt indetermination on our system design. To do this, two configurations are investigated : one with no Tip-tilt correction at all and one with a NGS Tip-tilt partial correction.

\subsection{No tilt correction : 100\% sky coverage}
\begin{figure}[h!]
   \begin{center}
   \begin{tabular}{c}
   \includegraphics[height=7cm]{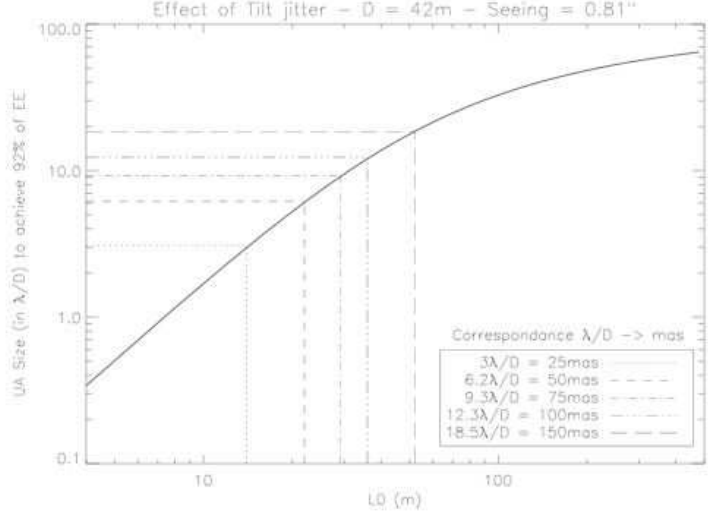}
   \end{tabular}
   \end{center}
   \caption[example] 
   { \label{TT}  Minimal size of the UA (expressed in $\lambda/D$ unit) required to concentrate 92\% of the energy versus $L_0$. Correspondences with typical Useful Aperture sizes expressed in 'mas' are also plot for comparison.}
   \end{figure}

As the wavefront Tilt cannot be measured from the LGS, we simply propose to not correct this mode. Doing this, the effect of Tilt variance will be to spread the PSF. In first approximation we will consider that the FWHM of a Tilt-uncorrected PSF is approximately given by $\sigma_{jitter}$, where $\sigma_{jitter}$ is defined by :
\begin{equation}
\label{eq1}
\sigma_{jitter} = \left(\frac{4}{2\pi}\frac{\lambda}{D}\right)\sigma_{Tilt}
\end{equation}
$\sigma_{Tilt}$ can be computed from general expressions as given by Conan et al\cite{Conan00}.\\
\\
If the UA size is larger than the FWHM of the Tilt-uncorrected PSF, a large proportion of the energy will stay in the aperture, even without Tip-tilt correction. Assuming a gaussian jitter much broader than the Airy peak, we will suppose that the final resulting PSF is a gaussian itself. From this, we can expect to concentrate approximately $92$\% of the energy if the squared UA size is equal to $2\sigma_{jitter}$. This condition can be written as :
\begin{equation}
\label{eq2}
 EE \simeq92\% \quad\mbox{if UA-Size}\ge 2\sigma_{jitter} = \frac{4}{\pi}\sigma_{Tilt}
\end{equation}
where, the UA Size and $\sigma_{jitter}$ are expressed in $\lambda/D$ unit.\\
\\
The value of $\sigma_{jitter}$ depends on seeing and also on the outer scale value $L_0$. This last parameter is of prime importance, especially when dealing with ELT. For this reason, we have computed the size of the UA required to concentrate $92$\% of the energy versus $L_0$. We consider the same turbulent profile as in section \ref{AOonAxis}, except for $L_0$ which value will vary. The telescope diameter is still D = $42$m. Results are presented on figure \ref{TT}.

Figure \ref{TT} shows that, for the seeing considered, and for $L_0$ $<$25m (resp. 50m) the performance criterion is met without Tilt correction for a UA size of 50mas $\simeq6.2\lambda/D$ (resp. 100mas $\simeq12.3\lambda/D$). These two examples demonstrate that the relevance of a Tip-Tilt free system is very dependent on the value and distribution of $L_0$. Of course the performance is also met for better seeings and smaller outer-scales. The probability to be in such domain (seeing and outer-scale better than given values) gives a lower bound of the useful observing time. This one can be deduced from the site turbulence statistics.  Assuming that the value of $L_0$ and Seeing are independent, this probability can be written as :
\begin{equation}
\label{obs}
P\left( Seeing \le S, L_0 \le L \right)=P\left(Seeing \le S \right)P \left(L_0 \le L \right)
\end{equation}
where, $S$ and $L$ are the maximal values of seeing and $L_0$ considered.\\
\\
The outer scale statistics has been measured by several campaigns\cite{Martin00}, showing a median value of $22$m and a log-normal distribution. Since for 50mas UA the limiting values for both seeing and outer-scale roughly correspond to median values, we can expect at least $25$\% of observing time.
For a system working with smaller UA, or in order to reach more observing time, a partial Tilt correction will be necessary.

\subsection{Partial Tip-Tilt correction}
\begin{figure}[h!]
   \begin{center}
   \begin{tabular}{cc}
      \includegraphics[height=5cm]{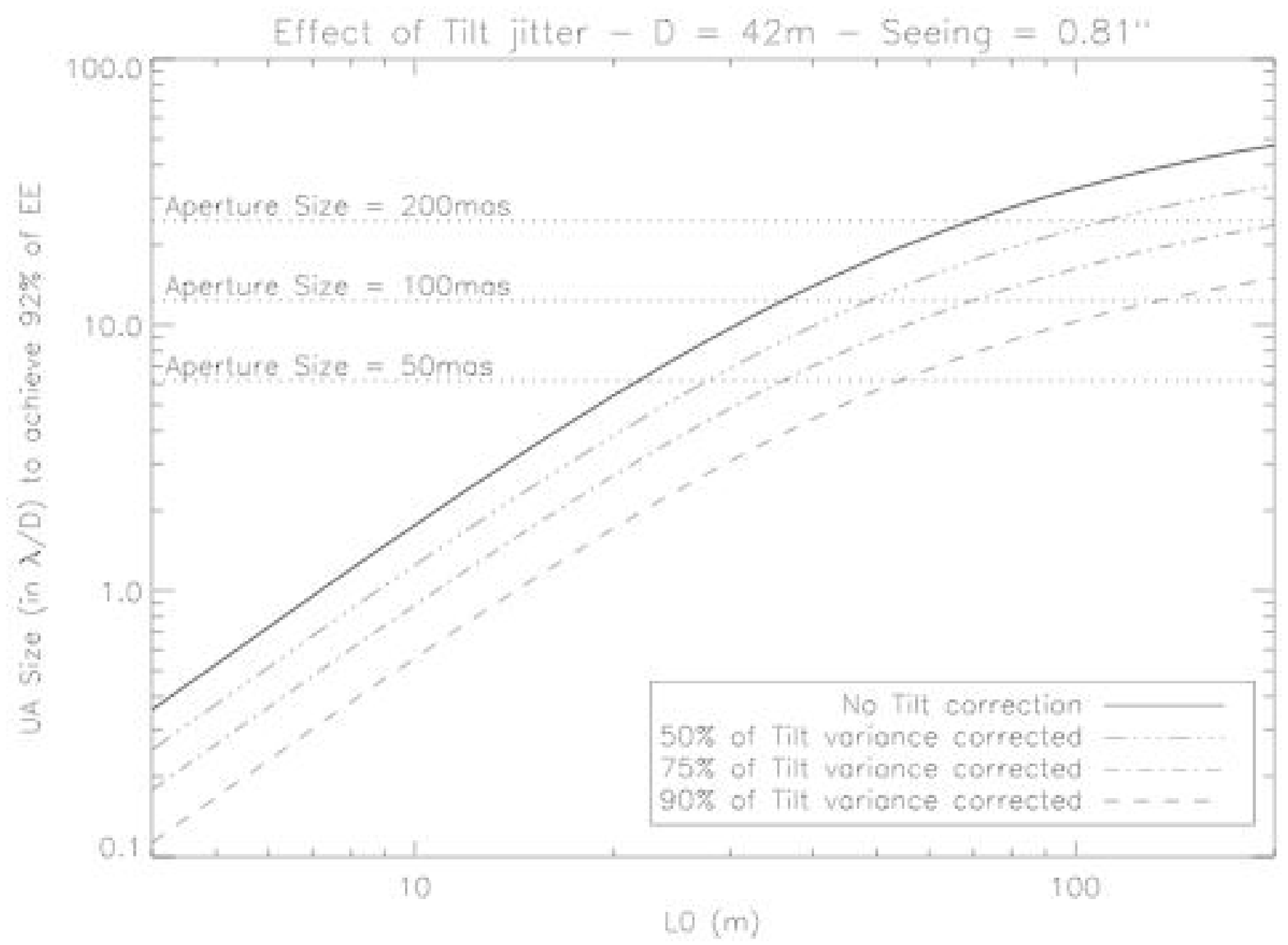} &
     \includegraphics[height=5cm]{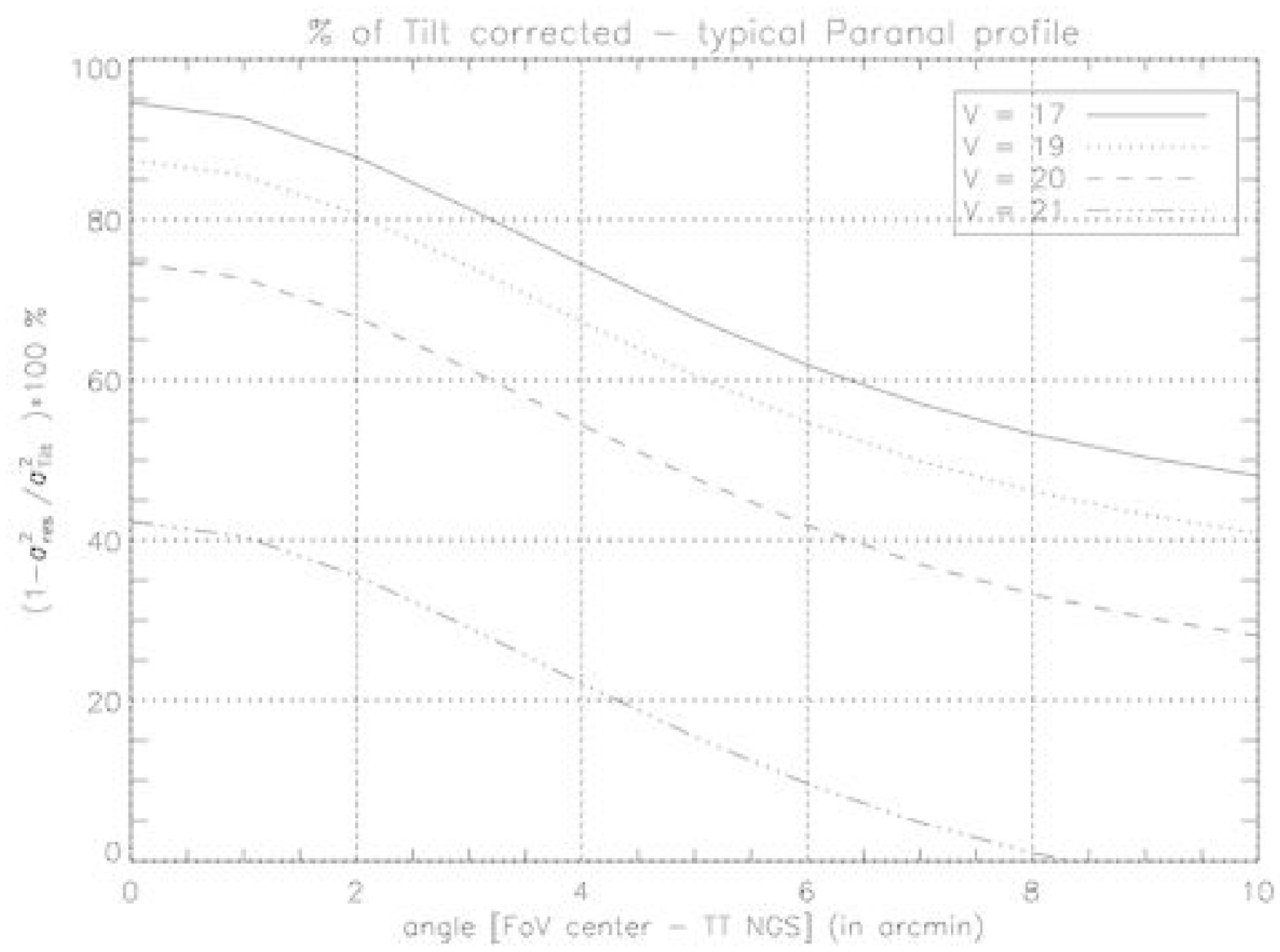} 
   \end{tabular}
   \end{center}
   \caption[example] 
   { \label{both}  \textit{Left : } Size of the UA required to concentrate $92\%$ of the energy in function of $L_0$ and for different percentage of Tilt variance correction. \textit{Right : } Correspondence between a given \% of Tilt variance correction and NGS configuration. We have consider the same Zero Point as in section \ref{mag}.\\}
   \end{figure}

Another solution to Tip-Tilt indetermination is to use two adaptive optics systems. One for the target, and one working on a NGS for the Tilt reference. For a given UA size, reducing the Tilt variance could relax the constraint on $L_0$ and so increase the achievable observing time.
To give an idea of the expected performance when Tilt variance is corrected, we keep the same condition as in eq. \ref{eq2}, but $\sigma_{Tilt}^2$ is replaced by $\sigma_{res}^{2}$ : the residual Tip-Tilt variance after correction by the NGS.
Figure \ref{both} (left) shows the UA size required to achieve $92$\% of EE when the Tilt variance is corrected at respectively 50, 75 and 90\%. For instance, correcting $75$\% of Tilt variance will provide a gain of a factor $2$ in terms of jitter. That approximately corresponds to a gain of a factor of $2$ in terms of aperture size : an aperture size around $60$mas $\simeq7.4\lambda/D$ could be large enough to concentrate the quasi totality of the energy up to $L_0$ = $40$m. This last condition is achieved approximately $40$\% of the time at Paranal\cite{Martin00}.
Using a NGS can thus brings a significant improvement in observing time.\\
However, using a NGS makes the issue of Sky Coverage to be again considered.
To do this, we determine what kind of NGS would be required in order to reduce the Tilt variance to an acceptable level (e.g. 50, 75 or 90\% of correction). We establish the link between the Tilt residual variance ($\sigma_{res}^{2}$) and observational parameters (NGS Magnitude and distance from optical axis, temporal bandwidth, ...) by splitting this variance into three terms as :
\begin{equation}
\label{eq3}
\sigma_{res}^{2} = \sigma_{pht}^{2}+\sigma_{temp}^{2}+\sigma_{aniso}^{2}
\end{equation}
The first term in Eq. \ref{eq3} represents the photon noise, the second term is the temporal error and the third one gives the contribution of anisoplanatism. Their expressions are given by :
\begin{eqnarray}
\label{eqq}
\sigma_{pht}^{2} = \frac{B_n}{f_t}\frac{\pi^2}{2ln2}\frac{1}{N_{ph}}\left(\frac{N_T}{N_{samp}}\right)^2\left(\frac{\lambda_{wfs}}{\lambda}\right)^2\\
\sigma_{temp}^{2} = 0.0162\left(\frac{v}{B_n D}\right)^2\left(\frac{D}{r_0}\right)^{5/3}\\
\sigma_{aniso}^{2} = \sum_{i=2,3}<(a_i(0)-a_i(\alpha))^2>
\end{eqnarray}
where $B_n$ is the bandwidth of the system, $f_t$ is the sampling frequency, $N_{ph}$ the number of photons received per aperture size and per integration time,  $N_T$ is the turbulent FWHM of the image (in pixels), $N_{samp}$ the diffraction FWHM of the image (in pixels), $\lambda$ the scientific wavelength, $v$ the wind speed and $\alpha$ the angular distance between the object and the reference source. The subscript $wfs$ refers to the WFS wavelength. \\
As we are dealing with low order correction, only few WFS sub-apertures are required. To measure the global tilt motion only one sub-aperture could be used. We only consider cases when our sub-aperture size ($d$) is larger than $r_{0_{wfs}}$. Under this condition, we can re-write eq. \ref{eqq} as :
\begin{equation}
\sigma_{pht}^{2} = \frac{B_n}{f_t}\frac{\pi^2}{2ln2}\frac{1}{N_{ph}}\left(\frac{d}{r_{0_{wfs}}}\right)^2\left(\frac{\lambda_{wfs}}{\lambda}\right)^2
\end{equation}
\\
Computing the tilt residual variance we express the percentage of correction ($1-\sigma_{res}^{2}/\sigma_{Tilt}^{2}$)*100\% versus the angular distance $\alpha$, the NGS magnitude and for different system design. In that way, Figure \ref{both} (right) shows the percentage of correction for a system working at $50$Hz with a single sub-aperture. For instance, to correct $75$\% of the Tilt variance, a NGS with V = $17$ at $\alpha$=$4'$ could be used. 

Then, we choose three NGS characteristics providing a correction of the Tilt variance of respectively $50$\%, $75$\% and $90$\%. These NGS are respectively describe by : V = $20$, $\alpha$=$4.5'$ ; V = $17$, $\alpha$=$4'$ and V = $17$, $\alpha$=$1.5'$.
We are now able to give an estimation of the SC when a partial Tilt correction is applied. Results are presented in Table \ref{sc}.\\
\begin{table}[h!]
\begin{center}
\begin{tabular}{|c|c|c|c|}
\hline  &  \bf{Galactic latitude = $30^\circ$} & \bf{Galactic Latitude = $60^\circ$} & \bf{Galactic Pole}\\ \hline \hline
V = $20$, $\alpha$=$4.5'$ & 100\% & 99.8\%  & 99.3\% \\ \hline
V = $17$, $\alpha$=$4'$ & 99.8\% & 87.3\% & 78.6\% \\ \hline
V = $17$, $\alpha$=$1.5'$ & 40.8\% & 16.5\% & 12.7\% \\ \hline
\end{tabular}
\caption{\label{sc} Sky coverage for different galactic latitudes and partial Tilt correction. The magnitude and angular distance of the NGS was optimized to maximize the achievable SC. SC are given in \% of the observable sky.}
\end{center}
\end{table}

For a good correction of Tilt variance (e.g. 75\%) we see that a quasi full sky coverage can be obtain even at high galactic latitudes. The solution of using a NGS to reduce the Tip-tilt variance can thus be implemented without any important loss in sky coverage.

\section{LGS MOAO}
\label{WFAO}

\begin{figure}[h!]
   \begin{center}
   \begin{tabular}{cc}
      \includegraphics[height=5cm]{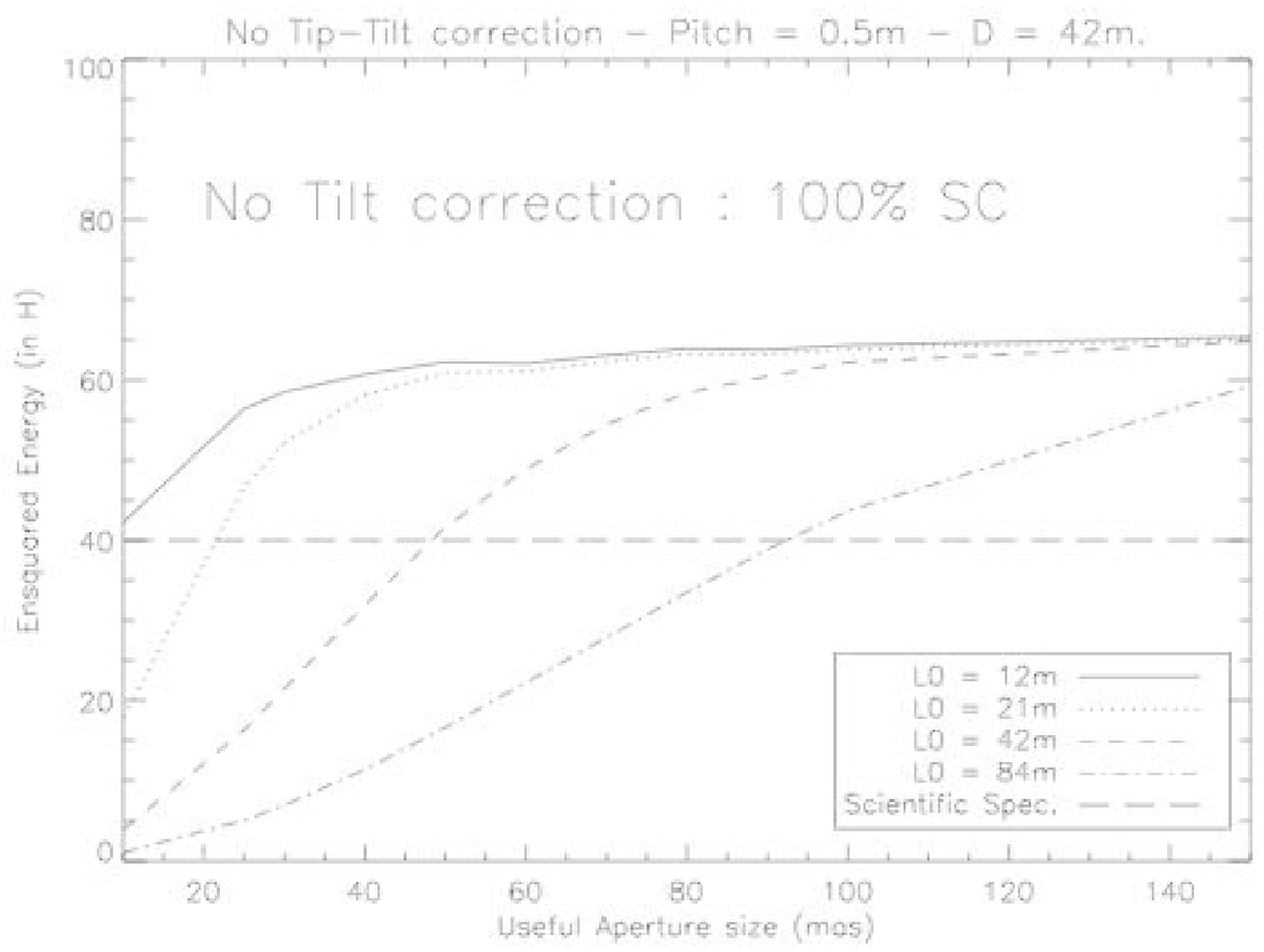} &
        \includegraphics[height=5cm]{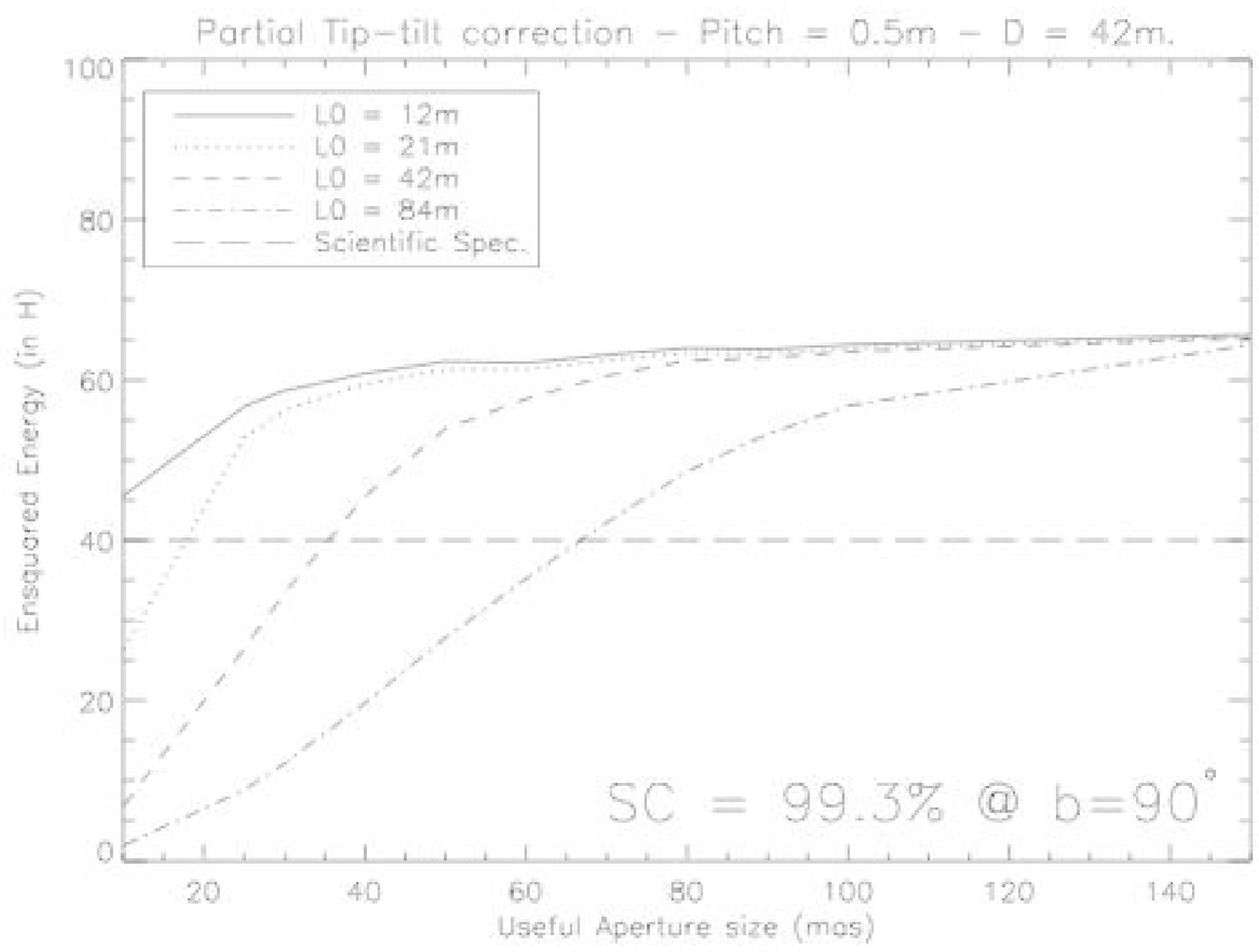} \\
           \includegraphics[height=5cm]{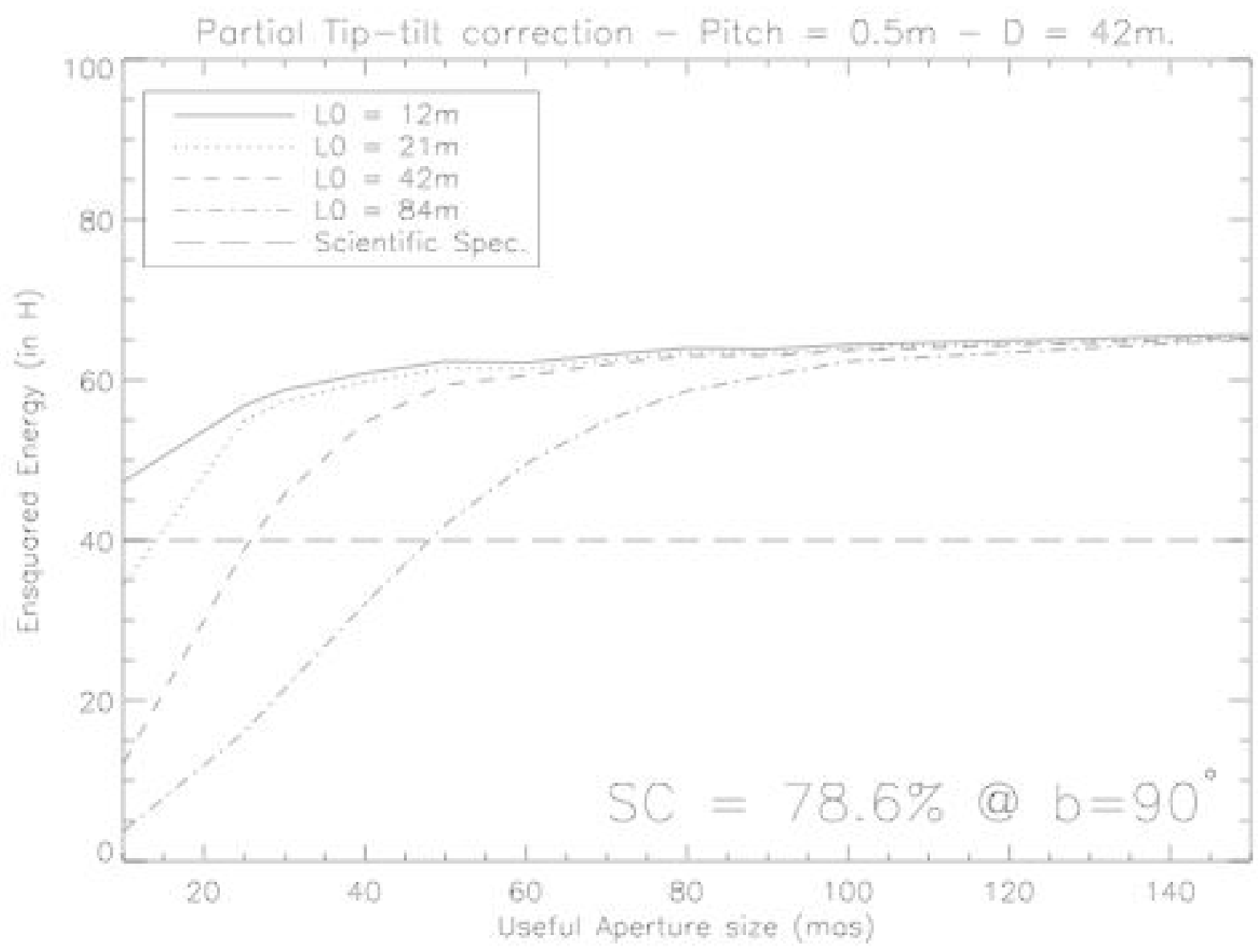} &
       \includegraphics[height=5cm]{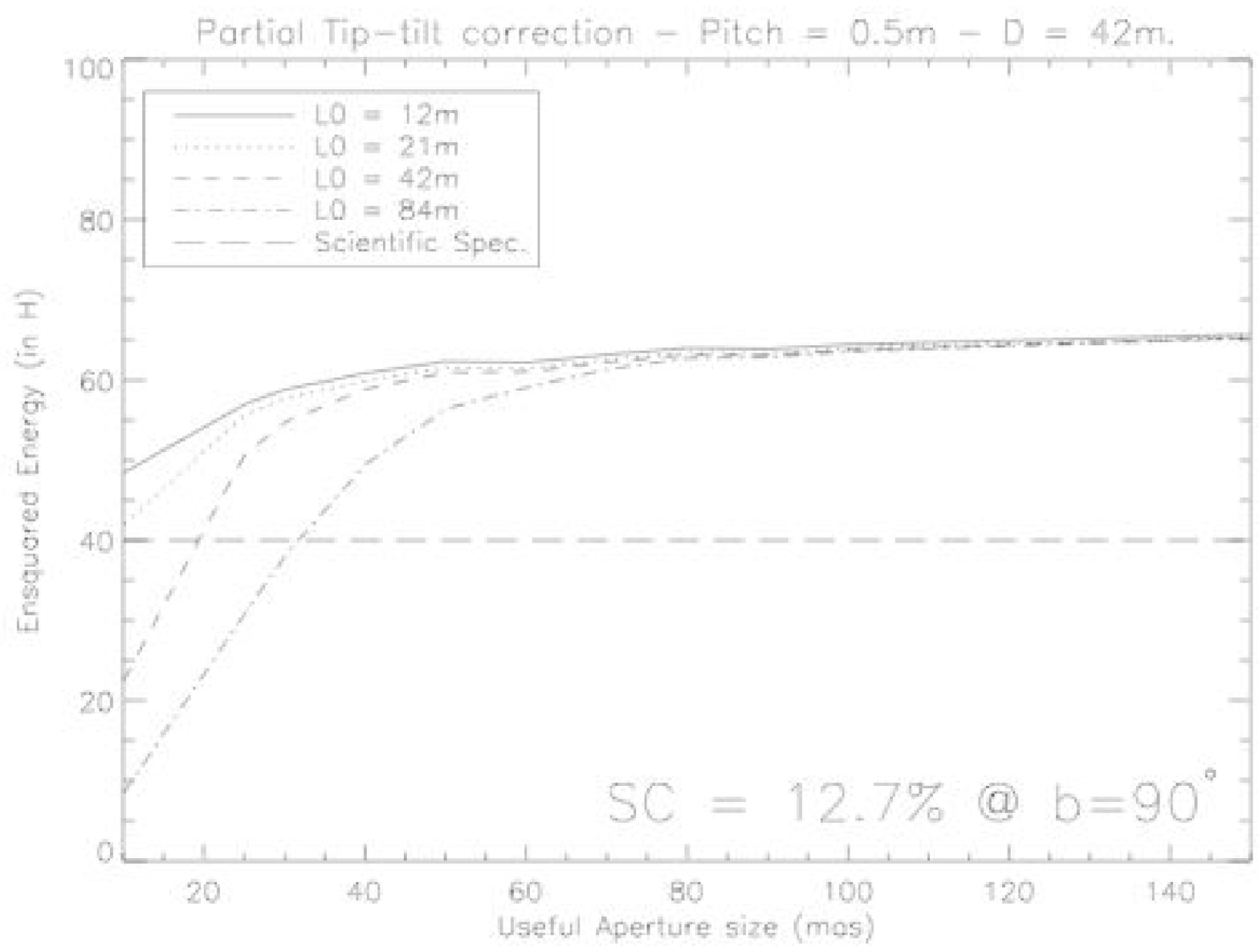}\\
      \multicolumn{2}{c}{\includegraphics[height=2.5cm]{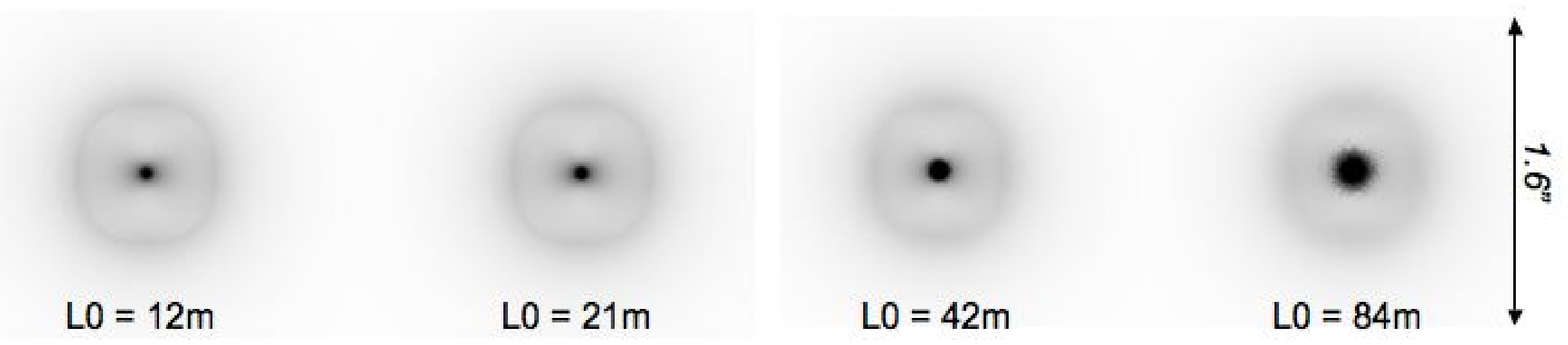}}
   \end{tabular}
   \end{center}
   \caption[example] 
   { \label{MOAO} LGS-WFS AO correction without or with partial correction of Tilt variance for MOAO concept. \textit{Top-Left : } No Tilt correction. \textit{Top-Right : } $50$\% of Tilt correction thanks to a NGS such as V = $20$ and $\alpha$=$4.5'$. \textit{Bottom-Left : } $75$\% of Tilt correction thanks to a NGS such as V = $17$ and $\alpha$=$4'$. \textit{Bottom-Right : } $90$\% of Tilt correction thanks to a NGS such as V = $17$ and $\alpha$=$1.5'$. \textit{Bottom : }AO-corrected PSFs for a different $L_0$ and no Tilt correction (Pitch = 0.5m - log sacle). }
   \end{figure}

The previous sections allow us to now address the design of an MOAO system.  Neglecting FA and tomographic issues we will assume that LGSs provide us with the turbulent phase information in each direction of interest, except for Tilt modes. The LGSs are bright enough to neglect WFS noise. Fitting, aliasing and temporal errors are taken into account as in Sect. \ref{doc}. Concerning the Tilt indetermination issue and following the approach presented in Sect. \ref{llgs}, we investigate four cases : one when no NGSs are available leading to total Tilt indetermination and three when a partial Tilt correction is provided by the NGSs described in table \ref{sc}. Results are presented in Figure \ref{MOAO}.

As in the previous section, Figure \ref{MOAO} shows that the achievable performance depends on the UA size, the Sky Coverage and $L_0$ (\% of observing time). With no Tilt correction at all, the influence of $L_0$ is important, and the largest UA must be considered : an UA size of at least $45$mas (resp. $90$mas) is needed to reach the scientific specifications, $40$\% (resp. $50$\%) of observing time. When a NGS is used, the influence of $L_0$ decreases at the cost of the SC. The graph associated to  $78.6\%$ of SC seems to correspond to an interesting compromise. In this configuration, the scientific specifications are met for a UA size of $50$mas and for  at least 50\% of observing time.\\
We see that, according to the most important parameter for scientific specifications it would be possible to find the best suited configuration. For instance we can imagine two instruments, one whose priority would be the Sky Coverage and one privileging the UA size. The first one will position itself towards a solution with no or few Tilt correction at the cost of the UA size, whereas the second one will allow a loss in SC in order to sufficiently correct the Tilt residual variance and concentrate a large proportion of the energy in small apertures. The best compromise will depend on the high level specifications. \\
Finally, when dealing with LGS, other low order modes (e.g. defocus, astigmatism) are corrupted. As the results for the Tilt mode are encouraging, we can easily imagine that the same approach for these modes could lead to interesting performance.

\section{Conclusion}
We have presented here some results concerning AO concepts for futures
ELTs and particularly for 3D spectroscopy.
We have shown that Sky Coverage issue will be a critical point when working with NGS-based system and the LGS solution seems inevitable for observations at high galactic latitudes. To solve the Tilt determination problem and to obtain a full Sky Coverage, we propose a new concept of rather simple Laser Guide Stars
systems without Tip-tilt correction. For median atmospheric conditions this concept fulfills the scientific requirements. For better performance, the use of a Tilt reference NGS even weak or far from optical axis reduces significantly the Tilt variance, and still allow a good Sky Coverage. Application to MOAO systems shows that thanks to partial Tilt correction, it could be possible to fulfill the scientific specifications in a large range of observational conditions while preserving a significant Sky Coverage.

\bibliography{report}   

\begin{thebibliography}{10}

\bibitem{Foy85}
R.~Foy and A.~Labeyrie, ``{Feasibility of adaptive telescope with laser
  probe},'' {\em A\&A}~{\bf 152}, pp.~L29--L31, nov~1985.

\bibitem{Hubin05}
N.~Hubin, R.~Arsenault, R.~Conzelmann, B.~Delabre, M.~Le~Louarn, S.~Stroebele,
  and R.~Stuik, ``{Ground Layer Adaptive Optics},'' in {\em C. R. Physique 6},
  2005.

\bibitem{Hammer01}
F.~Hammer, F.~Sayde, E.~Gendron, T.~Fusco, D.~Burgarella, V.~Cayatte, J.-M.
  Conan, F.~Courbin, H.~Flores, I.~Guinouard, L.~Jocou, A.~Lanon, G.~Monnet,
  M.~Mouchine, F.~Rigaud, D.~Rouan, G.~Rousset, V.~Buat, and Zamkotsian, ``{The
  FALCON concept: multi-object spectroscopy combined with MCAO in near-IR},''
  in {\em Scientifique Drivers for ESO Future VLT/VLTI Instrumentation},  ESO,
  (Garching Germany), june~2001.

\bibitem{Gendron05}
E.~Gendron, F.~Assemat, F.~Hammer, P.~Jagourel, F.~Chemla, P.~Laporte,
  M.~Puech, M.~Marteaud, F.~Zamkostian, A.~Liotard, J.-M. Conan, T.~Fusco, and
  N.~Hubin, ``{FALCON : Multi-Object OA},'' in {\em C. R. Physique 6},  2005.

\bibitem{LeLouarn98}
M.~{Le Louarn}, R.~{Foy}, N.~{Hubin}, and M.~{Tallon}, ``{Laser Guide Star for
  3.6- and 8-m telescopes: Performance and astrophysical implications},'' {\em
  MNRAS}~{\bf 295}, pp.~756--768, Apr.~1998.

\bibitem{Lelouarn02}
M.~{Le Louarn}, ``{Multi-Conjugate Adaptive Optics with laser guide stars:
  performance in the infrared and visible},'' {\em MNRAS}~{\bf 334},
  pp.~865--874, Aug.~2002.

\bibitem{Hook04}
I.~Hook, ``{Highlights from the science case for a 50- to 100-m extremely large
  telescope},'' in {\em Proceedings of the SPIE, Volume 5489, pp. 35-46
  (2004).},  J.~M. {Oschmann}, ed., pp.~35--46, Oct.~2004.

\bibitem{Rousselot00}
P.~{Rousselot}, C.~{Lidman}, J.-G. {Cuby}, G.~{Moreels}, and G.~{Monnet},
  ``{Night-sky spectral atlas of OH emission lines in the near-infrared},''
  {\em A\&A}~{\bf 354}, pp.~1134--1150, Feb.~2000.

\bibitem{Bouwens04}
R.~J. {Bouwens}, G.~D. {Illingworth}, J.~P. {Blakeslee}, T.~J. {Broadhurst},
  and M.~{Franx}, ``{Galaxy Size Evolution at High Redshift and Surface
  Brightness Selection Effects: Constraints from the Hubble Ultra Deep
  Field},'' {\em ApJ}~{\bf 611}, pp.~L1--L4, Aug.~2004.

\bibitem{Puech05}
M.~{Puech}, F.~{Hammer}, and H.~{Flores}, ``{Are luminous Compact Galaxies
  merger Remnants ?},'' {\em A\&A}~{\bf 105}, pp.~940--944, Sept.~2005.

\bibitem{Lehnert03}
M.~D. {Lehnert} and M.~{Bremer}, ``{Luminous Lyman Break Galaxies at z $\le$ 5
  and the Source of Reionization},'' {\em ApJ}~{\bf 593}, pp.~630--639,
  Aug.~2003.

\bibitem{Assemat06}
F.~{Assemat}, E.~{Gendron}, and F.~{Hammer}, ``{The FALCON concept :
  multi-object integral field spectroscopy combined with adaptive optics and
  atmospheric tomography. Principles and performances on a 8 meter
  telescope.},'' {\em MNRAS}~{\bf Submitted}, 2006.

\bibitem{Neichel05}
B.~{Neichel}, T.~{Fusco}, M.~{Puech}, J.-M. {Conan}, M.~{Le Louarn},
  E.~{Gendron}, F.~{Hammer}, G.~{Rousset}, P.~{Jagourel}, and P.~{Bouchet},
  ``{Adaptive optics concept for multi-object 3D spectroscopy on ELTs},'' in
  {\em IAU Symposium},  P.~{Whitelock}, M.~{Dennefeld}, and B.~{Leibundgut},
  eds., pp.~181--186, 2006.

\bibitem{Conan03}
R.~{Conan}, M.~{Le Louarn}, J.~{Braud}, E.~{Fedrigo}, and N.~{Hubin},
  ``{Results of AO simulations for ELTs},'' in {\em Future Giant Telescopes.
  Edited by Angel, J. Roger P.; Gilmozzi, Roberto. Proceedings of the SPIE,
  Volume 4840, pp. 393-403 (2003).},  J.~R.~P. {Angel} and R.~{Gilmozzi}, eds.,
  pp.~393--403, Jan.~2003.

\bibitem{Jolissaint06}
L.~{Jolissaint}, J.-P. {Veran}, and R.~{Conan}, ``{Analytical modeling of
  adaptive optics : foundations of the phase spatial power spectrum
  approach},'' {\em J. Opt. Soc. Am. A}~{\bf 23, No. 2}, feb~2006.

\bibitem{Fusco06}
T.~Fusco, A.~Blanc, M.~Nicolle, J.-L. Beuzit, V.~Michau, G.~Rousset, and
  N.~Hubin, ``{Sky coverage for MCAO systems : strategies and results},'' {\em
  MNRAS}~{\bf In press.}, 2006.

\bibitem{TF90}
M.~{Tallon} and R.~{Foy}, ``{Adaptive telescope with laser probe - Isoplanatism
  and cone effect},'' {\em A\&A}~{\bf 235}, pp.~549--557, Aug.~1990.

\bibitem{Viard02}
E.~{Viard}, M.~{Le Louarn}, and N.~{Hubin}, ``{Adaptive optics with four laser
  guide stars: correction of the cone effect in large telescopes},'' {\em
  ApOpt.}~{\bf 41}, pp.~11--20, Jan.~2002.

\bibitem{Pilkington87}
J.~D.~H. {Pilkington}, L.~{Thompson}, and C.~{Gardner}, ``{Artificial Guide
  Stars for Adaptive Imaging},'' {\em Nature}~{\bf 330}, pp.~116--+, Nov.~1987.

\bibitem{Foy95}
R.~{Foy}, A.~{Migus}, F.~{Biraben}, G.~{Grynberg}, P.~R. {McCullough}, and
  M.~{Tallon}, ``{The polychromatic artificial sodium star: a new concept for
  correcting the atmospheric tilt.},'' {\em A\&AS}~{\bf 111}, pp.~569--+,
  June~1995.

\bibitem{Ragazzoni96}
R.~{Ragazzoni}, ``{Absolute tip-tilt determination with laser beacons},'' {\em
  A\&A}~{\bf 305}, pp.~L13+, Jan.~1996.

\bibitem{Rigaut92}
F.~{Rigaut} and E.~{Gendron}, ``{Laser guide star in adaptive optics - The tilt
  determination problem},'' {\em A\&A}~{\bf 261}, pp.~677--684, Aug.~1992.

\bibitem{Conan00}
R.~{Conan}, J.~{Borgnino}, A.~{Ziad}, and F.~{Martin}, ``Analytical solution
  for the covariance and for the decorrelation time of the angle of arrival of
  a wave front corrugated by atmospheric turbulence,'' {\em OSAJ}~{\bf 17},
  pp.~1807--1818, Oct.~2000.

\bibitem{Martin00}
F.~{Martin}, R.~{Conan}, A.~{Tokovinin}, A.~{Ziad}, H.~{Trinquet},
  J.~{Borgnino}, A.~{Agabi}, and M.~{Sarazin}, ``{Optical parameters relevant
  for High Angular Resolution at Paranal from GSM instrument and surface layer
  contribution},'' {\em A\&AS}~{\bf 144}, pp.~39--44, May~2000.

\end{thebibliography}
\bibliographystyle{spiebib}   




\end{document}